\documentclass[preprint]{elsarticle}

\usepackage{graphicx}
\usepackage{amsmath}
\usepackage{amssymb}
\usepackage{float}

\let\originalleft\left
\let\originalright\right
\renewcommand{\left}{\mathopen{}\mathclose\bgroup\originalleft}
\renewcommand{\right}{\aftergroup\egroup\originalright}

\begin{document}

\title{Sinusoidal-signal detection by active, noisy oscillators on the brink of self-oscillation}

\author[ru,su]{D\'{a}ibhid \'{O} Maoil\'{e}idigh\corref{cor1}}
\ead{dmelody@stanford.edu}
\author[ru,hh]{A.~J.~Hudspeth}
\cortext[cor1]{Corresponding author}
\address[ru]{Laboratory of Sensory Neuroscience\\ The Rockefeller University\\
New York, NY 10065, USA\\
}
\address[su]{Present Address: Department of Otolaryngology--Head and Neck Surgery\\ Stanford University School of Medicine\\ Stanford, CA 94305, USA\\
}
\address[hh]{Howard Hughes Medical Institute\\ The Rockefeller University\\
New York, NY 10065, USA\\
}

\date{\today}

\begin{abstract}
\noindent
Determining the conditions under which an active system best detects sinusoidal signals is important for numerous fields. It is known that a quiescent, deterministic system possessing a supercritical Hopf bifurcation is more sensitive to sinusoidal stimuli the closer it operates to the bifurcation. To understand signal detection in many natural settings, however, noise must be taken into account. We study the Fokker-Planck equation describing the \emph{sinusoidally forced} dynamics of a noisy supercritical or subcritical Hopf oscillator. To distinguish an oscillator's motion owing to sinusoidal forcing from that provoked by noise, we employ the phase-locked amplitude and vector strength, which are zero in the absence of an external signal. The phase-locked amplitude and entrainment to frequency-detuned forcing---but not resonant forcing---peak as functions of the control parameter. These peaks occur near but not at the bifurcations. Moreover, an oscillator can detect stimuli over the broadest frequency range when it spontaneously oscillates near a Hopf bifurcation. Although noise exerts the greatest effect on the phase-locked amplitude when a Hopf oscillator is near a Hopf bifurcation, the oscillator nevertheless performs best as a sinusoidal-signal detector when it operates close to the bifurcation. The oscillator's ability to differentiate detuned signals from noise is greatest with it autonomously oscillates near to but not at the bifurcation.
\end{abstract}


\maketitle

\section{Introduction\label{intro}}
The performance of a sinusoidal-signal detecter is enhanced if it operates quiescently in the proximity of a supercritical Hopf bifurcation \cite{choe98,eguiluz00,camalet00}. The response to sinusoidal forcing of a quiescent deterministic system operating near a supercritical Hopf bifurcation is frequency-selective and nonlinear, with the response magnitude and the sharpness of frequency tuning increasing as the bifurcation is approached \cite{choe98}. The range over which the input is nonlinearly compressed also rises as the system nears the bifurcation \cite{eguiluz00, camalet00}. To determine whether operation near a Hopf bifurcation is useful for signal detection, however, we must take into consideration the effects of noise, operation on the oscillatory side of the bifurcation, and subcritical as well as supercritical Hopf bifurcations.

Bifurcations in noisy systems often do not coincide with those in the corresponding deterministic systems \cite{arnold03}. Indeed, noise can create new bifurcations and destroy or change deterministic ones. For concreteness, we can study systems possessing bifurcations in the deterministic sense, to which noise has been added. We consider the normal forms of supercritical and subcritical Hopf oscillators in the presence of noise, which define noisy Hopf oscillators.

The effects of sinusoidal forcing on noisy oscillators have been described in many contexts. Stochastic oscillations, stochastic bifurcations, stochastic resonance, stochastic synchronization, and many other phenomena have been revealed through the study of specific systems such as van der Pol oscillators, integrate-and-fire models, and various phase oscillators \cite{lindner04,pikovsky03}. Often the response to sinusoidal forcing of a particular type of oscillator operating near a Hopf bifurcation has been investigated. For example, the FitzHugh-Nagumo model has been studied near a supercritical Hopf bifurcation whereas the behavior of the Noyes-Field-Thomson model has been discussed near a subcritical Hopf bifurcation \cite{longtin93,pei95,hohmann96}.

Several authors have studied the behavior of noisy Hopf oscillators in the absence of deterministic forcing. The stationary Fokker-Planck equation corresponding to the normal form can be solved exactly for complex, additive, Gaussian white noise with uncorrelated components \cite{hempstead67,ushakov05} or with weakly correlated components in the supercritical case \cite{baras82}, or for particular types of multiplicative noise \cite{graham82a,zheng85,triana08}. Indeed, the power spectrum for an isochronous oscillator can be calculated from the Fokker-Planck equation and is approximately Lorentzian when the oscillator is quiescent, in which case there is a closed-form approximation for the spectral width, and when the system spontaneously oscillates far from the bifurcation \cite{hempstead67,ushakov05}. The power spectrum of a non-isochronous, supercritical Hopf oscillator can be strongly non-Lorentzian, however, but can be expressed as an infinite sum of Lorentzian's \cite{gleeson06}. In general, noise in the normal form of a random dynamical system is multiplicative, colored, and cross-correlated \cite{coullet85,arnold03}.

To understand the basic behavior of a sinusoidally forced noisy Hopf oscillator, one can consider the simplest type of noise. Complex Gaussian white noise, which possesses uncorrelated real and imaginary components, can be added to the sinusoidally-forced Hopf normal form \cite{stratonovich67b}. If the amplitude of oscillation is approximately constant, the phase dynamics is equivalent to that of a Brownian particle in an inclined potential and the degree of entrainment can be found analytically \cite{haken67}. More generally, the response function of isochronous, self-oscillating Hopf oscillators to sinusoidal forcing at or near the natural frequency is constant for weak forcing and declines compressively for stimuli of increasing amplitude, with an exponent for frequency-tuned forcing and weak noise that can be calculated analytically \cite{lindner09}. A self-oscillatory system exhibits strong compression far from a supercritical Hopf bifurcation or at any distance from a subcritical bifurcation. Moreover, for the supercritical Hopf normal form with a stimulus frequency close to the natural frequency and this kind of noise, the linear response amplitude possesses the same analytical form as the deterministic, quiescent case with renormalized parameters \cite{julicher09}. We lack a more complete understanding, however, of a noisy Hopf oscillator's response as a function of the control parameter.

To determine the behaviors of noisy sinusoidal-signal detectors that are inherent to Hopf bifurcations, we study the response to sinusoidal forcing of supercritical and subcritical Hopf oscillators as a function of the control parameter. We discuss specific cyclostationary solutions to the Fokker-Planck equation describing a noisy Hopf oscillator. These solutions vary in time at the frequency of driving and describe the system's behavior after transients have vanished. We find the response of the Hopf oscillators for both low and high noise levels and forcing amplitudes, and limit the forcing amplitude only to avoid the possibility of chaotic dynamics at very high levels of forcing \cite{perez82,rajasekar88}. 

The correspondence between a system operating near a Hopf bifurcation and a Hopf oscillator may break down for sufficiently great forcing and noise levels. Nonetheless, the response of Hopf oscillators to strong forcing or their response in the presence of a substantial amount of noise serves to illustrate basic characteristics that we expect to be common for a large class of oscillators. To determine if a noisy Hopf oscillator's ability to detect a sinusoidal signal is augmented when the system operates near the bifurcation, we focus on the phase-locked amplitude, the response function, and the vector strength. The phase-locked amplitude and response function quantify the oscillator's entrained response to the stimulus, whereas the vector strength quantifies the degree of entrainment.

\section{Noisy Hopf Oscillators\label{Hopf}}
A Hopf bifurcation occurs when a fixed point changes stability at a critical value of a control parameter \cite{kuznetsov,guckenheimer83,strogatz}. The dynamics of any system that possesses a Hopf bifurcation is two-dimensional in the vicinity of the critical point and can be described by the complex variable $z \equiv z_\mathrm{R} + i z_\mathrm{I}$. To linear order in $z$,
\begin{eqnarray}
\mathbf{x} &=& \mathbf{x^*} + \mathbf{c_\mathrm{R}}z_\mathrm{R} + \mathbf{c_\mathrm{I}}z_\mathrm{I}\\
&=& \mathbf{x^*} + \left[\mathbf{c_\mathrm{R}}\cos(\phi) + \mathbf{c_\mathrm{I}}\sin(\phi)\right]|z| \mathrm{,}
\end{eqnarray}
in which $\mathbf{x}$ is the state vector of the original system, $\mathbf{x^*}$ is its value at the fixed point corresponding to the bifurcation, and $\mathbf{c_\mathrm{R}}$ and $\mathbf{c_\mathrm{I}}$ are constant real vectors of the same dimension as $\mathbf{x}$ that depend on the system's particular structure and parameters \cite{ipsen98}. The angle $\phi$ satisfies $\tan\phi = z_\mathrm{I}/z_\mathrm{R}$. The dynamics of any component of the original system $x_\mathrm{i}$ is qualitatively similar to the dynamics of $z_\mathrm{R}$, $z_\mathrm{I}$, or $|z|$, which are described by the truncated normal form for a Hopf bifurcation.

We study the Hopf normal form in the presence of sinusoidal forcing with additive white noise possessing uncorrelated real and imaginary components:
\begin{equation}
\label{HopfNF}
\dot{z} = (\mu+i\omega_0)z+(b+ib')|z|^2 z+(c+ic')|z|^4 z+fe^{i(\omega t+\theta)}+\eta \mathrm{,}
\end{equation}
in which $\mu$ is the control parameter, $\omega_0$ is the natural frequency, and $b$, $b'$, $c$, and $c'$ are coefficients defining the system's nonlinearity. The complex forcing term $F(t)=fe^{i(\omega t+\theta)}$ has amplitude $f$, frequency $\omega$, and phase $\theta$. The noise $\eta(t)$ is complex and white and satisfies $\langle \eta(t) \rangle=0$, $\langle\eta(t)\eta(t')\rangle = 0$, and $\langle\eta(t)\eta^\mathrm{\dagger}(t')\rangle = 4d\delta(t-t')$, in which $\langle\rangle$ represents the ensemble average, $\eta^\mathrm{\dagger}(t)$ is the complex conjugate of $\eta(t)$, and $d$ is the strength of the noise \cite{stratonovich67b,gleeson06,julicher09}.

The correspondence between the Hopf oscillator defined by Eq.~\ref{HopfNF} and a system operating near a Hopf bifurcation generally relies on the forcing being weak and the noise being small, additive, and white \cite{julicher09,zhang11}. We can, however, also view Eq.~\ref{HopfNF} as defining the dynamics of a specific system possessing a Hopf bifurcation, namely, a Hopf oscillator.

In the absence of forcing and noise, a Hopf bifurcation occurs when $\mu = 0$; the system oscillates when $\mu>0$ and possesses a stable fixed point for $\mu<0$. When $b<0$ and $c=0$ the amplitude of oscillation grows continuously from zero at the bifurcation, corresponding to a supercritical Hopf bifurcation. Oscillations of nonzero amplitude exist at a subcritical Hopf bifurcation that occurs when $b>0$ and $c<0$. The oscillations are generated by a saddle-node of limit cycles bifurcation at $\mu = b^2/4c < 0$. For $b^2/4c<\mu<0$ a stable fixed point is encircled by an unstable limit cycle, which is in turn surrounded by a stable limit cycle. We refer to the situations with $\mu>0$ as the oscillatory sides of the bifurcations, the supercritical case of $\mu<0$ and subcritical case of $\mu<b^2/4c$ as the quiescent sides of the bifurcations, and the subcritical situation in which $b^2/4c<\mu<0$ as the coexistence region.

Equation \ref{HopfNF} can be rewritten in the rotating frame of the forcing as
\begin{equation}
\label{HopfNFrot}
\dot{y} = (\mu-i\delta\omega)y+(b+ib')|y|^2 y+(c+ic')|y|^4 y+f+\eta e^{-i(\omega t+\theta)} \mathrm{,}
\end{equation}
in which $y \equiv ze^{-i(\omega t+\theta)}$ and $\delta\omega \equiv \omega-\omega_0$ is the frequency detuning. Because we focus on the response of the oscillator relative to the forcing, the factor $e^{-i(\omega t+\theta)}$ has no effect on the results. In the case of complex Gaussian noise, for which the real and imaginary components are Gaussian distributed with variance $2d$, the probability density $P(y,t)$ for $y$ satisfies the corresponding Fokker-Planck equation:
\begin{eqnarray}
\label{FProt}
\partial_\mathrm{t} P
&=&-\partial_\mathrm{R}\left[\left(\mu y_\mathrm{R}+\delta\omega y_\mathrm{I}+(by_\mathrm{R}-b'y_\mathrm{I})\rho^2+(cy_\mathrm{R}-c'y_\mathrm{I})\rho^4+f\right)P-d\partial_\mathrm{R}P\right]\nonumber\\
&&-\partial_\mathrm{I}\left[\left(\mu y_\mathrm{I}-\delta\omega y_\mathrm{R}+(by_\mathrm{I}+b'y_\mathrm{R})\rho^2+(cy_\mathrm{I}+c'y_\mathrm{R})\rho^4\right)P-d\partial_\mathrm{I}P\right]\mathrm{,}
\end{eqnarray}
in which $y \equiv y_\mathrm{R} + iy_\mathrm{I}$, $\partial_\mathrm{R}\equiv\partial/\partial y_\mathrm{R}$, $\partial_\mathrm{I}\equiv\partial/\partial y_\mathrm{I}$, and the radial coordinate $\rho(y_\mathrm{R},y_\mathrm{I}) \equiv \sqrt{y_\mathrm{R}^2+y_\mathrm{I}^2}$ \cite{stratonovich67b,gardiner}. Owing to the additive nature of the noise, the Stratonovich and It\^o forms of the Fokker-Plank equation are identical. A special case of Eq.~\ref{FProt} corresponding to the unforced, isochronous, supercritical Hopf oscillator ($f = \delta \omega = b' = c = c' = 0$) has been discussed previously \cite{hempstead67}.
 
The parameters $b'$ and $c'$ define a nonlinear relationship between the frequency of spontaneous oscillations and the control parameter. In the supercritical case the response to sinusoidal forcing when $b'$ is nonzero can be multivalued and the forcing generates a large number of bifurcations \cite{zhang11,wiser15}. To capture the most basic effects of noise on a forced Hopf oscillator, we restrict further analysis to the case of $b'=0$ and $c'=0$, which defines an isochronous Hopf oscillator \cite{pikovsky03}.

The solution for the steady-state distribution $P_\mathrm{s}(y)$ of $y$ in the case of zero detuning, $\delta\omega = 0$, is given by
\begin{equation}
\label{sdist}
P_\mathrm{s} = N\mathrm{exp}\left(\frac{\mu\rho^2}{2d}+\frac{b\rho^4}{4d}+\frac{c\rho^6}{6d}+\frac{fy_\mathrm{R}}{d}\right)\mathrm{,}
\end{equation}
in which $N$ is a normalization constant, as reported previously for the supercritical case \cite{stratonovich67b}. Although no closed-form solution for Eq.~\ref{FProt} is apparent when the detuning $\delta\omega$ is nonzero, we can find the steady-state distribution numerically using the exact solution for zero detuning to set the initial conditions and to approximate the boundary conditions (Eq.~\ref{sdist}, \ref{numericalFPApp}).

To find the average response, we consider the Fourier amplitude of $z(t)$ at the frequency $\omega_\mathrm{n} = 2\pi n/T$
\begin{equation}
\label{FT}
\tilde{z}(\omega_\mathrm{n})\equiv\frac{1}{T}\int_{-\frac{T}{2}}^{\frac{T}{2}} e^{-i\omega_\mathrm{n}t}z(t)\,dt\mathrm{,}
\end{equation}
in which $T$ is the observation time for $z(t)$ and $n$ is an integer.
As $T \to \infty$, the average Fourier amplitude $\langle\tilde{z}(\omega')\rangle_\mathrm{zs}$ is nonzero only if $\omega'=\omega$, in which $\langle \rangle_\mathrm{zs}$ denotes the long-time ensemble average (\ref{chimagApp}).
The magnitude of the average response amplitude at the frequency of driving, which we call the phase-locked amplitude, is then defined as
\begin{equation}
\label{PLres}
R \equiv |\langle \tilde{z}(\omega) \rangle_\mathrm{zs}| = |\langle y\rangle_\mathrm{s}|\mathrm{,}
\end{equation}
in which $|\langle \tilde{z}(\omega) \rangle_\mathrm{zs}|$ is finite for all values of $T$ including $T \to \infty$ and the right-hand side is a consequence of $P_\mathrm{s}$ being stationary (\ref{chimagApp}). The symbol $\langle \rangle_\mathrm{s}$ denotes the stationary ensemble average in the rotating frame.
The magnitude of the corresponding response function, also known as the sensitivity, is
\begin{equation}
\label{chimag}
|\tilde{\chi}(\omega)| \equiv \frac{|\langle \tilde{z}(\omega) \rangle_\mathrm{zs}|}{f}=\frac{|\langle y\rangle_\mathrm{s}|}{f}\mathrm{.}
\end{equation}
The degree of entrainment is quantified by the vector strength
\begin{equation}
\label{VS}
V \equiv \left|\lim_{T\to\infty} \frac{1}{T}\int_0^\mathrm{T} e^{i\psi(t)}\,dt\right| = \left|\langle e^{i\psi}\rangle_\mathrm{s}\right|\mathrm{,}
\end{equation}
in which $\psi \equiv \phi - \omega t - \theta$ is the phase difference between the response and driving.
The right-hand side of Eq.~\ref{VS} is based on the assumption that the system is ergodic in the frame that rotates at the frequency of driving. In the presence of noise, the phase-locked amplitude and the vector strength are nonzero only if the forcing amplitude $f\neq0$.

The phase-locked amplitude and vector strength signify two different aspects of a system's response to sinusoidal driving. The phase-locked amplitude increases as entrainment to the stimulus grows, but can rise without limit after perfect entrainment has been achieved. Perfect entrainment is equivalent to a constant phase difference between the stimulus and the response. The vector strength is a bounded measure of entrainment, in which a value of unity indicates perfect entrainment and a value of zero implies none. The phase-locked amplitude and vector strength can be calculated numerically using the closed-form expression for the steady-state probability density (Eq.~\ref{sdist}) when the forcing is tuned, or from the numerical solution to Eq.~\ref{FProt} when the detuning is nonzero.

We note that we can also calculate the average of the amplitude's magnitude at the frequency of driving $\langle|y|\rangle_\mathrm{s} = \langle\rho\rangle_\mathrm{s} \geq |\langle y\rangle_\mathrm{s}|$ from the probability density. More generally, higher-order moments of the amplitude's magnitude can be derived systematically from only three functions (\ref{amplmomApp}). For an unforced supercritical Hopf oscillator, analytical expressions for the moments can be derived iteratively. Because all moments of the amplitude's magnitude can be nonzero in the absence of forcing, we focus in the remainder of this manuscript on the phase-locked amplitude and the vector strength.
\begin{figure}[t]
\centering
\includegraphics{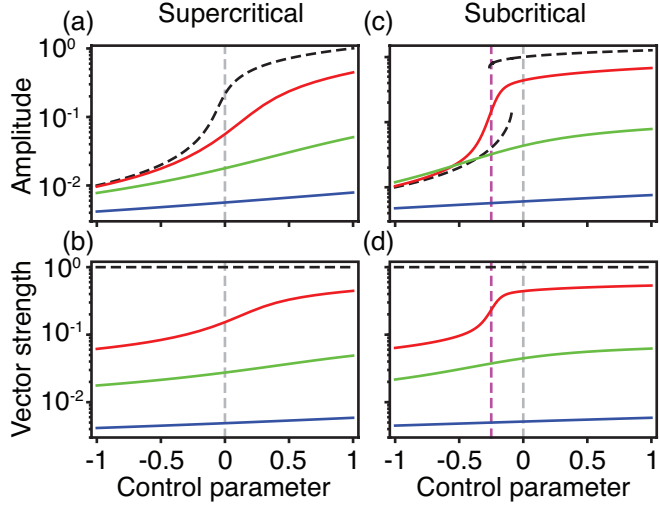}
\caption{The phase-locked amplitude and vector strength are shown as functions of the control parameter $\mu$ at the natural frequency near supercritical (a, b) and subcritical (c, d) Hopf bifurcations. The noise level is $d=0$ (black, dashed), $d=0.01$ (red), $d=0.1$ (green), or $d=1$ (blue). In the deterministic subcritical case in panel (c), two stable responses lines exist for a limited range of control-parameter values. The Hopf bifurcations occur at $\mu=0$ (gray, dashed) and a saddle-node of limit cycles bifurcation occurs at $\mu =  b^2/4c = -0.25$ (magenta, dashed) in the subcritical case. The forcing amplitude is $f=0.01$. Forcing has little effect on the phase-locked amplitude for large control parameter values, for which the amplitude increases as $\mu^{1/2}$ in the supercritical case and as $\mu^{1/4}$ for a subcritical oscillator. For all figures in this manuscript the natural frequency $\omega_0 = 1$, $b = -1$ and $c = 0$ in the supercritical case, or $b = 1$ and $c = -1$ in the subcritical case. \label{chiVSvsmu_fig}}
\end{figure}
\subsection{Sinusoidal-signal detection as a function of the control parameter}
\subsubsection{For tuned forcing, the phase-locked amplitude and vector strength grow monotonically as the control parameter increases.}
In the presence or absence of noise, the phase-locked amplitude of a resonantly forced Hopf oscillator grows as the control parameter increases (Fig.~\ref{chiVSvsmu_fig}). The closer the system operates to the Hopf bifurcation, however, the greater is the influence of noise. Owing to the rise in the amplitude of spontaneous oscillations as the control parameter grows, the effects of noise diminish. Noise likewise does not affect the phase-locked amplitude significantly in the limit of very negative values of the control parameter, for which the response to both deterministic and stochastic input is small.

In the subcritical case, a single averaged response replaces the two stable responses seen in the deterministic limit when $9 b^2/20 c < \mu < 0$ and $f_\mathrm{1}(\mu,b,c) < f < f_\mathrm{2}(\mu,b,c)$ (\ref{perfectlockApp}). For some noise levels, the phase-locked amplitude is greater than the deterministic response, but this enhancement depends on the control parameter's value.

In the absence of noise, the system is perfectly entrained by the stimulus for all values of the control parameter. In contrast to the deterministic limit, the vector strength of a noisy oscillator rises as the control parameter is increased. Noise thus introduces a qualitative change in the response of the system to sinusoidal stimuli.

For tuned forcing, $\langle y_\mathrm{R}\rangle_\mathrm{s} > 0$ and $\langle y_\mathrm{I}\rangle_\mathrm{s} = 0$ such that $|\langle y \rangle_\mathrm{s}| = \langle y_\mathrm{R} \rangle_\mathrm{s}$. Thus we find
\begin{eqnarray}
\frac{\partial |\langle y \rangle_\mathrm{s}|}{\partial \mu} &=& \frac{\partial \langle y_\mathrm{R}\rangle_\mathrm{s}}{\partial \mu}\nonumber\\
&=& \int^\infty_{-\infty} \int^\infty_{-\infty} \frac{\partial P_\mathrm{s}}{\partial \mu} y_\mathrm{R} dy_\mathrm{R} dy_\mathrm{I}\nonumber\\
&=& \frac{\langle \rho^2 y_\mathrm{R} \rangle_\mathrm{s} - \langle \rho^2 \rangle_\mathrm{s}\langle y_\mathrm{R} \rangle_\mathrm{s}}{2d}\nonumber\\
&=& \frac{\mathrm{cov}(y_\mathrm{R}, \rho^2)}{2 d}\mathrm{,}
\end{eqnarray}
in which $\mathrm{cov}()$ is the covariance under the distribution $P_\mathrm{s}$. It can be shown similarly that
\begin{eqnarray}
\frac{\partial |\langle e^{i\psi} \rangle_\mathrm{s}|}{\partial \mu} &=& \frac{\mathrm{cov}(y_\mathrm{R}/\rho, \rho^2)}{2 d}\mathrm{.}
\end{eqnarray}
The rise in the responsiveness measures as a function of $\mu$ is a consequence of the specific form of the distribution $P_\mathrm{s}$, for which $\mathrm{cov}(y_\mathrm{R}, \rho^2) > 0$ and $\mathrm{cov}(y_\mathrm{R}/\rho, \rho^2) > 0$.

Because the phase-locked amplitude and the vector strength grow as the control parameter is increased, a resonantly forced, noisy, self-oscillating Hopf oscillator performs better as a detector of sinusoidal forces the \emph{further} it operates from a Hopf bifurcation. The oscillator may be required to detect a range of stimulus frequencies, however, leading us to examine its response to detuned forcing.

\subsubsection{For detuned forcing, the phase-locked amplitude and the stochastic system's vector strength peak near a Hopf bifurcation.}
\begin{figure*}[t]
\centering
\includegraphics[width=\textwidth]{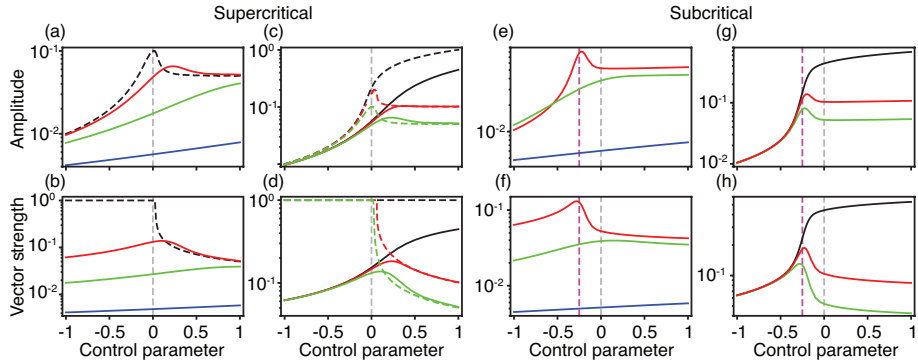}
\caption{The phase-locked amplitude and vector strength are shown as functions of the control parameter for detuned forcing at $f=0.01$. (a, b, e, f) The response for forcing at a frequency below the natural frequency $\omega=0.9\,\omega_0$ is show for a noise level of $d = 0$ (dashed black), $d = 0.01$ (red), $d=0.1$ (green), or $d= 1$ (blue). (c, d, g, h) The response in the presence of noise ($d=0.01$) is shown for different amounts of detuning $|\delta\omega| = 0$ (black), $|\delta\omega| = 0.05\,\omega_0$ (red), or $|\delta\omega| = 0.1\,\omega_0$ (green). (c, d) The supercritical response in the absence of noise is shown for the detuning $|\delta\omega| = 0$ (black, dashed), $|\delta\omega| = 0.05\,\omega_0$ (red, dashed), or $|\delta\omega| = 0.1\,\omega_0$ (green, dashed). In the subcritical case, Hopf bifurcations occur at $\mu=0$ (gray, dashed) and a saddle-node of limit cycles bifurcation occurs at $\mu =  b^2/4c$ (magenta, dashed).
\label{chiVSvsmu_detuned_fig}}
\end{figure*}

A Hopf oscillator responds to detuned forcing in a qualitatively different manner than it does to tuned forcing. In the presence of noise, the phase-locked amplitude and the vector strength peak as functions of the control parameter for all non-zero levels of noise and detuning (Fig.~\ref{chiVSvsmu_detuned_fig}). For both types of bifurcation, the peaks move to larger values of the control parameter $\mu$ and their magnitudes decrease as the noise level increases. Owing to the limited range of $\mu$ shown, it may appear as if the phase-locked amplitude and vector strength plateau for large values of the control parameter and that they do not peak at finite value of $\mu$. We argue, however, that $R \to 0$ and $V \to 0$ as $\mu \to \pm \infty$ (\ref{CPlimApp}). Consequently, the responsiveness measures for detuned forcing must peak at finite values of the control parameter.

The mechanism that creates the peak in the phase-locked amplitude is clearest for a supercritical Hopf bifurcation without noise. Starting on the quiescent side of the bifurcation, the system is perfectly entrained by the stimulus and, owing to increasing proximity to the bifurcation, the phase-locked amplitude grows as the control parameter is raised. The response continues to rise with the control parameter after the system passes through the Hopf bifurcation until entrainment declines owing to a Hopf or saddle-node bifurcation of the forced system (\ref{perfectlockApp}). The phase-locked amplitude thereafter decreases as detuned forcing fails to entrain the oscillator. It is increasingly difficult to entrain an oscillator to a detuned stimulus as the amplitude of spontaneous motion rises. As for the tuned case, the phase-locked amplitude is affected by noise only when the system is close to a supercritical Hopf bifurcation.

For the deterministic case, the vector strength drops monotonically once the control parameter exceeds a critical value. In contrast, the vector strength peaks in the presence of noise owing to its decline for increasingly negative values of the control parameter. The oscillator's response to sinusoidal forcing decreases and is consequently more corrupted by noise. We also note that the maximum in the vector strength occurs at a value of the control parameter different from that at which the phase-locked amplitude peaks.

In the subcritical case, the stochastic phase-locked amplitude and vector strength peak near the saddle node of limit cycles bifurcation that is associated with the subcritical Hopf bifurcation.

Finally, the positions and sizes of the peaks for the phase-locked amplitude and vector strength depend only on the magnitude of the detuning and not on its sign (Fig.~\ref{chiVSvsmu_detuned_fig}). The phase-locked amplitude and vector strength for detuned forcing asymptotically approach the tuned limit, however, as the control parameter declines.

\begin{figure*}
\centering
\includegraphics[width=\textwidth]{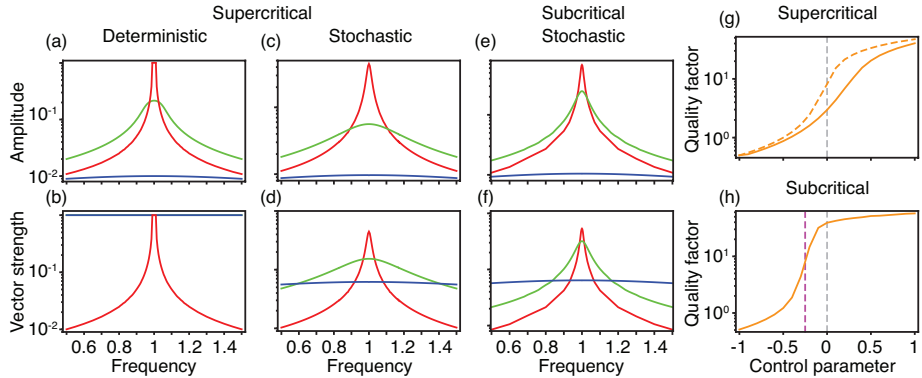}
\caption{(a-f) The phase-locked amplitude and vector strength are shown as functions of the driving frequency for a force of $f=0.01$ and values of the control parameter $\mu = 1$ (red), $\mu = 0$ (green), or $\mu = -1$ (blue). (a, b) The deterministic supercritical case. (c, d) The stochastic supercritical case. (e, f) The stochastic subcritical case. (g) The quality factor is shown as a function of control parameter for the supercritical case with (solid line) and without (dashed line) noise. The Hopf bifurcation occurs at $\mu=0$ (gray, dashed). (h) The quality factor is shown as a function of the control parameter for the subcritical bifurcation in the presence of noise. A Hopf bifurcation occurs at $\mu=0$ (gray, dashed) and a saddle-node of limit cycles bifurcation occurs at $\mu =  b^2/4c$ (magenta, dashed). In all panels, the noise level for the stochastic cases is $d=0.01$.
\label{chiVSvsw_fig}}
\end{figure*}
\subsection{Phase-locked amplitude and entrainment as functions of the stimulus frequency}
The phase-locked amplitude and vector strength depend on the driving frequency for supercritical and subcritical Hopf oscillators (Fig.~\ref{chiVSvsw_fig}). In the absence of noise, the phase-locked amplitude of a supercritical system displays a maximum as a function of frequency for all values of the control parameter. In contrast, the vector strength is unity for all frequencies on the quiescent side of the bifurcation, but is frequency-tuned on the oscillatory side. In the presence of noise, however, both quantities peak at the natural frequency for all values of $\mu$.

There is a qualitative difference between frequency tuning on the oscillatory and quiescent sides of a Hopf bifurcation. For a quiescent oscillator, the phase-locked amplitude increases at all stimulus frequencies the closer the oscillator is to the bifurcation. In contrast, the phase-locked amplitude of a self-oscillating system at stimulus frequencies far from the resonant frequency \emph{decreases} as the control parameter increases. This difference is preserved when noise is taken into account.

The sharpness of frequency tuning can be quantified by calculating the quality factor $Q \equiv \omega_\mathrm{max}/\Delta\omega$, in which $\omega_\mathrm{max}$ is the frequency at which the phase-locked amplitude peaks at $R_\mathrm{max}$ and $\Delta\omega$ is the width of the frequency range over which $R>R_\mathrm{max}/\sqrt{2}$. The larger the quality factor, the greater the frequency selectivity. With or without noise, the sharpness of frequency tuning grows as the control parameter $\mu$ increases. In the supercritical case, noise decreases the frequency selectivity at all control-parameter values. The effects of noise are greatest, however, near the Hopf bifurcation. Frequency tuning approaches the deterministic limit for very positive or negative values of the control parameter.

\begin{figure}[t]
\centering
\includegraphics{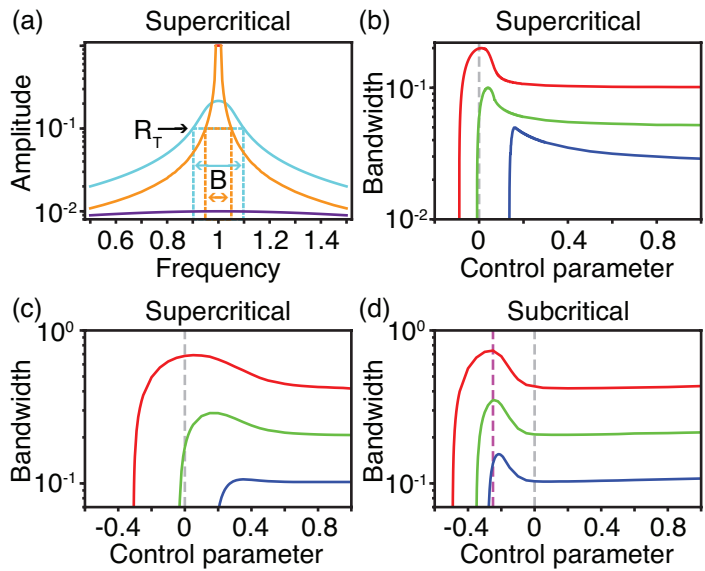}
\caption{The threshold bandwidth peaks as a function of the control parameter. (a) The phase-locked amplitude for a deterministic supercritical Hopf bifurcation depends on the stimulus frequency for a control-parameter value of $\mu = 1$ (orange), $\mu = 0$ (cyan), or $\mu = -1$ (purple). The threshold bandwidth $B$ is the range of frequencies for which the phase-locked amplitude exceeds a threshold $R_\mathrm{T}$. The amplitude does not exceed the threshold of 0.1 for $\mu = -1$ (purple), but the threshold bandwidth is smaller when the control parameter is 1 (orange) than when it is 0 (cyan). (b) The threshold bandwidth in the absence of noise peaks near a supercritical Hopf bifurcation (gray, dashed) for a threshold of 0.1 (red), 0.2 (green), or 0.4 (blue). (c, d) The threshold bandwidth peaks near a Hopf bifurcation (gray, dashed) when the noise level $d=0.01$ for a threshold of 0.025 (red), 0.05 (green), or 0.1 (blue). (d) A saddle node of limit cycles bifurcation occurs at $\mu =  b^2/4c$ (magenta, dashed). The stimulus force $f=0.01$ for all panels.
\label{BWmu_fig}}
\end{figure}

\subsection{Maximal threshold bandwidth near a Hopf bifurcation.}
As we have seen, a Hopf oscillator responds best to sinusoidal stimuli with frequencies near its resonant frequency. The range of stimulus frequencies for which the stimulus is detected depends on the control parameter's value. We suppose that an oscillator detects a stimulus if the magnitude of its phase-locked amplitude equals or exceeds a threshold $R \geq R_\mathrm{T}$. We define the \textit{threshold bandwidth} $B$ to be the size of the frequency range for which detection occurs, that is, $B \equiv |\{\omega: R \geq R_\mathrm{T}\}|$ (Fig.~\ref{BWmu_fig}a).

In the presence or absence of noise, the threshold bandwidth is greatest when a supercritical Hopf system operates close to and on the oscillatory side of the bifurcation (Fig.~\ref{BWmu_fig}). As the threshold increases, the bandwidth declines and the bandwidth's peak moves to larger values of the control parameter. When we compare the bandwidth for a specific threshold in the absence of noise to that in the presence of noise, we see that adding noise decreases the size of the peak and shifts it to greater values of the control parameter. The threshold bandwidth eventually falls to zero for sufficiently positive or negative control parameters.

The threshold bandwidth of a stochastic subcritical Hopf system depends on the control parameter in a manner similar to that of a system possessing a supercritical Hopf bifurcation. The main difference is that the bandwidth peak occurs near the saddle node of limit cycles bifurcation, that is, at the boundary of the region possessing spontaneous oscillations. On the side of the peak closer to the oscillatory region, the threshold bandwidth initially rises slowly as the control parameter increases further. The bandwidth finally falls to zero, however, as the system moves far from the bifurcation in either the quiescent or the oscillatory direction.
\begin{figure}[H]
\centering
\includegraphics{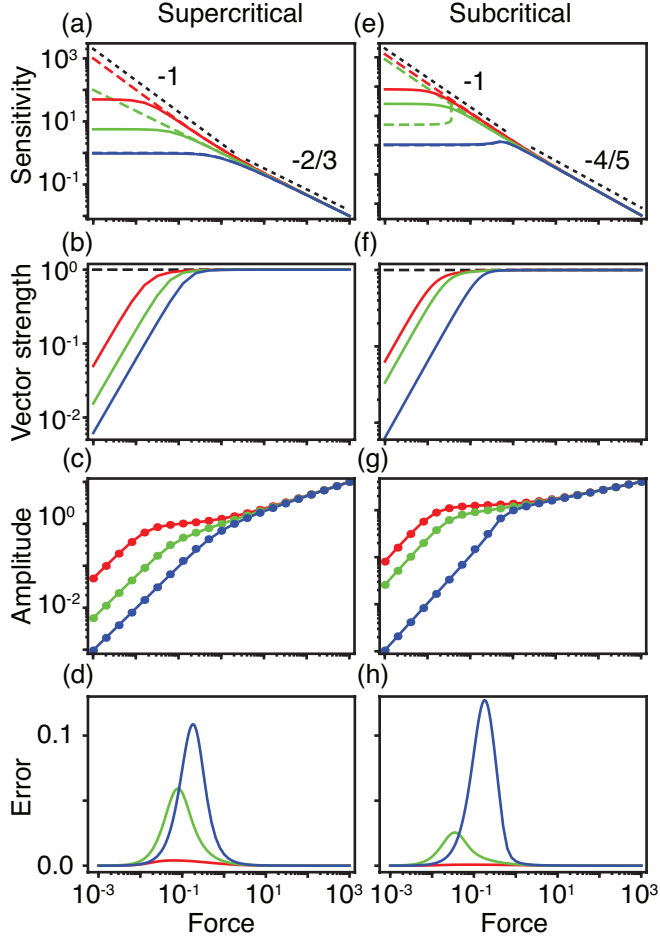}
\caption{The response function, vector strength, phase-locked amplitude, and an approximation's relative error for forcing at the natural frequency near supercritical (a--d) and subcritical (e--h) bifurcations are shown as functions of the forcing amplitude. Stable responses are depicted in the absence of noise (dashed lines) and in the presence of noise ($d = 0.01$, solid lines).  (a, b, c, d) The supercritical control parameter is $\mu=1$ (red), $\mu=0$ (green), or $\mu=-1$ (blue). (e, f, g, h) The subcritical control parameter is $\mu=1$ (red), $\mu=-0.2$ (green), or $\mu=-1$ (blue). (a, e) The black dotted lines labeled by the number $\alpha$ indicate the range of forces corresponding to the power law $|\tilde{\chi}|\sim f^\alpha$. (e) In the absence of noise, two stable responses exist for low forcing amplitudes (green dashed lines). (b, f) Because the vector strength is unity for all forcing amplitudes and control-parameter values in the deterministic cases, it is shown as a black dashed line. (c, g) The phase-locked amplitude (solid lines) and an approximation based on the vector-strength (Eqs.~\ref{chilinandVSlin} and \ref{chiandVSinfty}, solid points) are shown. (d, h) The error of the approximation to the phase-locked amplitude relative to the exact calculation is illustrated (Eq.~\ref{yfromVSerror}).
\label{chiVSvsf_fig}}
\end{figure}
In all cases, the threshold bandwidth is very sensitive to changes in the control parameter on the side of the peak closer to the quiescent region. This sensitivity arises from the failure of the system to achieve threshold for any stimulus frequency when it operates on the quiescent side of and far from a Hopf bifurcation. The gradual drop on the oscillatory side of the peak reflects the loss of entrainment that occurs as the system's spontaneous oscillations grow in amplitude. To ensure a large threshold bandwidth while maintaining robustness of the bandwidth to parameter changes, it is consequently best to position the operating point of a Hopf system near to but on the larger-$\mu$ side of the bandwidth's peak. In summation, the range of frequencies that an oscillator can detect is maximized when it spontaneously oscillates near a Hopf bifurcation.

\subsection{Responsiveness as a function of the forcing amplitude}
\subsubsection{Noise allows an oscillator to detect small tuned forces.}
Many of the phenomena we have discussed thus far rely on the forcing amplitude being small. We now examine how sinusoidal-signal detection depends on the signal's amplitude. The response function captures the sensitivity of an oscillator to a change in the forcing amplitude (Eq.~\ref{chimag}). For tuned sinusoidal forcing with $f \gg \sqrt{2d}$, the system's responsiveness depends little on the control parameter and is similar to the deterministic limit (Fig.~\ref{chiVSvsf_fig}). The vector strength is close to unity and the response function decreases with a power law characteristic of the bifurcation type, $|\tilde{\chi}|\sim f^{-2/3}$ in the supercritical case and $|\tilde{\chi}|\sim f^{-4/5}$ in the subcritical case \cite{eguiluz00,camalet00,lindner09}. Far from the bifurcation, noise has little effect on the response function of a quiescent oscillator, but the vector strength grows to unity as the forcing rises.

On the oscillatory sides of the bifurcations, noise reduces the response function and the vector strength and changes the power law for weak forces from $|\tilde{\chi}|\sim f^{-1}$ to $|\tilde{\chi}|\sim f^{0}$. This linear-response regime is associated with a decline in entrainment owing to noise and is coincident with a range of forces for which the vector strength increases linearly. In the subcritical case, the two deterministic responses observed for $9 b^2/20 c < \mu < 0$ and $f_\mathrm{1}(\mu,b,c) < f < f_\mathrm{2}(\mu,b,c)$ are replaced by a single averaged response (\ref{perfectlockApp}). Our calculations agree with analytical results for autonomously oscillating systems based on small fluctuations in $\rho$ owing to noise, in which case the compressive exponents can also be calculated analytically \cite{lindner09}. For weakly forced quiescent oscillators, however, noisy fluctuations in $\rho$ cannot be neglected, but we find the sensitivity's dependence on the forcing amplitude to be similar to that on the oscillatory sides of the bifurcations.

The decrease in the phase-locked amplitude with greater forcing allows the oscillator to compress many orders of magnitude in input into a significantly smaller output range. This compressive region extends to smaller forces as the control parameter increases. In the deterministic case, compression following the power law $|\tilde{\chi}|\sim f^{-1}$ is not useful for signal detection, for the oscillator does not respond to the stimulus. In the stochastic instance, however, weak forces can be discriminated, because the vector strength and the phase-locked amplitude increase linearly with the forcing amplitude.

Analytical expressions for the response function and the vector strength on either side of the bifurcation can be found for weak forcing by expanding the exact solution for the steady-state distribution (Eq.~\ref{sdist}) to linear order in $f$, yielding
\begin{equation}
\label{chilin}
|\tilde{\chi}_\mathrm{l}| \equiv \frac{\left|\langle y_\mathrm{R}y\rangle_\mathrm{s0}\right|}{d}
\end{equation}
and
\begin{equation}
\label{VSlin}
V_\mathrm{l} \equiv \frac{\left|\langle y_\mathrm{R}e^{i\psi}\rangle_\mathrm{s0}\right|f}{d}\mathrm{,}
\end{equation}
in which $\langle\rangle_\mathrm{s0}$ represents averaging over the steady-state ensemble in the absence of forcing (\ref{chimagApp}). Equation \ref{chilin} relates the linear response $\tilde{\chi}_\mathrm{l}$ to the covariance of the unstimulated system $\langle y_\mathrm{R}y\rangle_\mathrm{s0}$ and is essentially a fluctuation-dissipation relation for Hopf oscillators \cite{landau_lifshitzStatPhys}. This expression accords with previous estimates that either assume that the noise has little effect on $\rho$ or specialize to the supercritical bifurcation \cite{lindner09,julicher09}. 

The expressions for the linear response (Eqs.~\ref{chilin} and \ref{VSlin}) can be used to relate the magnitude of the phase-locked amplitude for weak forcing $|\langle y\rangle_\mathrm{sl}|$ to $V_\mathrm{l}$ as
\begin{equation}
\label{chilinandVSlin}
|\langle y\rangle_\mathrm{sl}| = \frac{\left|\langle y_\mathrm{R}y\rangle_\mathrm{s0}\right|V_\mathrm{l}}{\left|\langle y_\mathrm{R}e^{i\psi}\rangle_\mathrm{s0}\right|}\mathrm{.}
\end{equation}
For large forcing a similar relationship holds between the phase-locked amplitude and the vector strength because the system's response becomes essentially deterministic. From Eq.~\ref{VS},
\begin{eqnarray}
\label{chiandVSinfty}
\left|\langle y\rangle_\mathrm{s\infty}\right|
&=& \frac{\left|\langle y\rangle_\mathrm{s\infty}\right|\left |\langle y_\mathrm{R}\rangle_\mathrm{s\infty}\right|V_\mathrm{\infty}}{\left|\langle y_\mathrm{R}\rangle_\mathrm{s\infty}\right|\left|\langle e^{i\psi}\rangle_\mathrm{s\infty}\right|}\nonumber\\
&=& \frac{\left|\langle y_\mathrm{R} y \rangle_\mathrm{s\infty}\right|V_\mathrm{\infty}}{\left|\langle y_\mathrm{R} e^{i\psi}\rangle_\mathrm{s\infty}\right|} \mathrm{,}
\end{eqnarray}
in which the subscript $\infty$ denotes averaging over the steady-state ensemble in the limit of arbitrarily large forcing. An approximation for $R$ is thus
\begin{equation}
\label{RfromV}
R_\mathrm{V} = \frac{\left|\langle y_\mathrm{R}y\rangle_\mathrm{s}\right|V}{\left|\langle y_\mathrm{R}e^{i\psi}\rangle_\mathrm{s}\right|}\mathrm{.}
\end{equation}
The magnitude of the phase-locked amplitude agrees with $R_\mathrm{V}$ independent of the value of the control parameter for small (Eq.~\ref{chilinandVSlin}) and large (Eq.~\ref{chiandVSinfty}) forcing, but the agreement is worse for intermediate forces at which the approximation based on $V$ systematically exceeds the true phase-locked amplitude (Fig.~\ref{chiVSvsf_fig}).
The relative error is given by
\begin{eqnarray}
\label{yfromVSerror}
\frac{R_\mathrm{V}-R}{R}\mathrm{,}
\end{eqnarray}
whose peak rises as the control parameter declines. We thus encounter an unusual situation in which an approximation is best in the limits of small and large inputs, but is least accurate for intermediate values. The relation between $R$ and $V$ can be expressed formally, however, as
\begin{eqnarray}
R &=&\frac{\left\langle \rho I_\mathrm{1} \left(\frac{f\rho}{d}\right) \right\rangle_\mathrm{s0} V}{\left\langle I_\mathrm{1} \left(\frac{f\rho}{d}\right) \right\rangle_\mathrm{s0}}\mathrm{,}
\label{PLvVSBessel}
\end{eqnarray}
in which $I_\mathrm{n}(x)$ is the modified Bessel function of the first kind of integer order $n$ (\ref{chimagApp}).

\begin{figure}[t]
\centering
\includegraphics{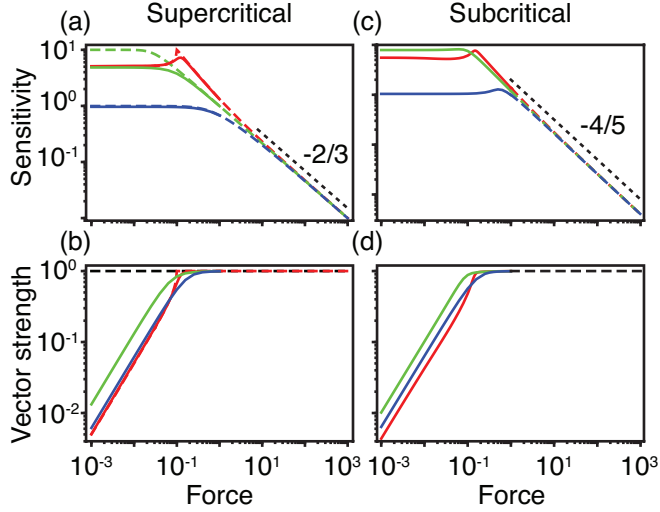}
\caption{The response function and vector strength are shown as functions of the forcing amplitude at a frequency below the natural frequency, $\omega=0.9\,\omega_0$, near supercritical (a, b) and subcritical (c, d) bifurcations. The stochastic curves ($d = 0.01$, solid lines) are plotted only for the forcing levels $f\leq1$ because they are indistinguishable from the deterministic curves (dashed lines) for greater forcing. (a, b) The supercritical control parameter is $\mu=1$ (red), $\mu=0$ (green), or $\mu=-1$ (blue). The deterministic vector strength is shown using a black dashed line when $\mu=0$ and $\mu=-1$ because it is unity for all forces. (c, d) The subcritical control parameter is $\mu=1$ (red), $\mu=-0.2$ (green), or $\mu=-1$ (blue). The deterministic lines are shown only for $f\geq1$; two stable responses exist for lower levels of forcing. (d) For $f\geq1$, the vector strength is essentially unity (black dashed). (a, c) The black dotted lines labeled by the number $\alpha$ indicate the range of forces corresponding to the power law $|\tilde{\chi}|\sim f^\alpha$.
\label{chiVSvsf_detuned_fig}}
\end{figure}

\subsubsection{Small detuned forces are best detected near a Hopf bifurcation.}
When the forcing level is large, the responses to frequency-detuned forcing are similar to those to tuned forcing (Fig.~\ref{chiVSvsf_detuned_fig}). There are, however, several qualitative differences for weak forcing. On the oscillatory side of a deterministic supercritical Hopf bifurcation, the phase-locked amplitude can possess a peak as a function of forcing level owing to a decrease in entrainment at a critical forcing amplitude (\ref{perfectlockApp}). In addition, the system possesses a linear response on the oscillatory side of the bifurcations in both the deterministic and stochastic cases. With or without noise, the vector strength increases linearly with the forcing amplitude on the oscillatory side of the bifurcation and is larger when the system operates near a Hopf bifurcation than when it operates further away. The deterministic and stochastic systems are consequently most sensitive to small forces when they operate near a Hopf bifurcation.

\begin{figure}[t]
\centering
\includegraphics{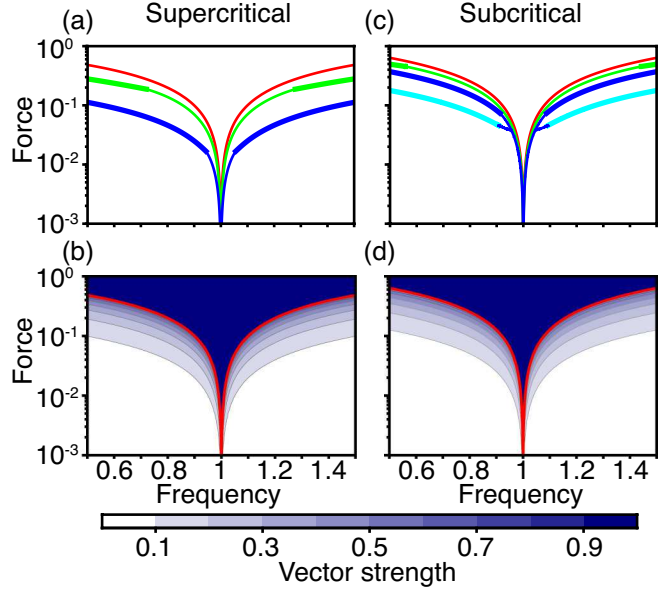}
\caption{State diagrams of the driven deterministic system as a function of the forcing amplitude and frequency. (a, c) Wedge-shaped regions of perfect entrainment are bounded below by bifurcation lines of the forced system for different control-parameter values. Thick lines denote Hopf bifurcations and thin lines correspond to saddle-node bifurcations. The  full set of bifurcation curves is not shown \cite{pikovsky03}. (a) The value of the control parameter is $\mu = 1$ (red), $\mu = 0.5$ (green), or $\mu = 0.1$ (blue). Thick lines represent supercritical Hopf bifurcations. (c) The control-parameter value is $\mu = 1$ (red), $\mu = 0.1$ (green), or $\mu = -0.2$ (blue, cyan). When $\mu = 1$ or $\mu = 0.1$, thick lines demarcate supercritical Hopf  bifurcations. When $\mu = -0.2$, the system is imperfectly locked in the belts delimited by contours of supercritical Hopf (blue thick), subcritical Hopf (cyan thick), and saddle-node bifurcations (blue thin). At this value of the control parameter, a perfectly phase-locked amplitude coexists with an imperfectly locked response for the regions demarcated above by saddle-node bifurcation lines and lines of subcritical Hopf bifurcations. (b, d) The vector strength is shown as a function of the stimulus amplitude and frequency in the case of $\mu = 1$. The analytical boundaries for perfect phase locking (red) bound the area for which $V = 1$.\label{ArnoldTongueDet_fig}}
\end{figure}
\subsection{Entrainment as a function of the stimulus amplitude and frequency\label{tongue_sect}}
Quiescent deterministic systems can be perfectly entrained by sinusoidal forcing of any amplitude and frequency. In contrast, self-oscillatory systems exhibit perfect entrainment for a limited range of stimulus amplitudes and frequencies illustrated by state diagrams of the forced system (Fig.~\ref{ArnoldTongueDet_fig} and \ref{perfectlockApp}). The region of perfect entrainment can be calculated analytically and is demarcated by lines corresponding to bifurcations of the forced system. For large forces, the width of this wedge-shaped region is less than that of the one-to-one Arnold tongue over which the phase difference $\psi$ between the system's response and stimulus is bounded \cite{pikovsky03}. On the oscillatory sides of the Hopf bifurcations, the wedge is delimited by lines of saddle-node bifurcations for small forcing amplitudes and detuning, and by lines of supercritical Hopf bifurcations for large forcing and detuning.

For a set of control-parameter values near a subcritical Hopf bifurcation, there can be two stable responses for low forcing amplitudes and sufficiently small detuning. In the case of large detuning, a stable response can coexist with a stable limit cycle, a situation that corresponds to less-than-perfect entrainment. Depending on the initial conditions, $\psi$ is either constant, corresponding to $V=1$, or time-dependent, in which case $V<1$. For both types of Hopf oscillator, however, the width of the perfect-entrainment sector decreases as the control parameter grows, yielding sharper frequency selectivity at all forcing amplitudes.

Outside the area of perfect entrainment for a self-oscillating system, the vector strength decreases as the forcing falls or the detuning rises. Wedge-shaped regions in which entrainment exceeds a threshold can still be defined, however, by contours of constant vector strength (Fig.~\ref{ArnoldTongueDet_fig}). These wedges asymptotically collapse onto the region of perfect entrainment as the forcing amplitude declines to zero. In the coexistence region near a subcritical Hopf oscillator, the contours of constant vector strength depend on the choice of initial conditions.
\begin{figure}
\centering
\includegraphics[height=6in]{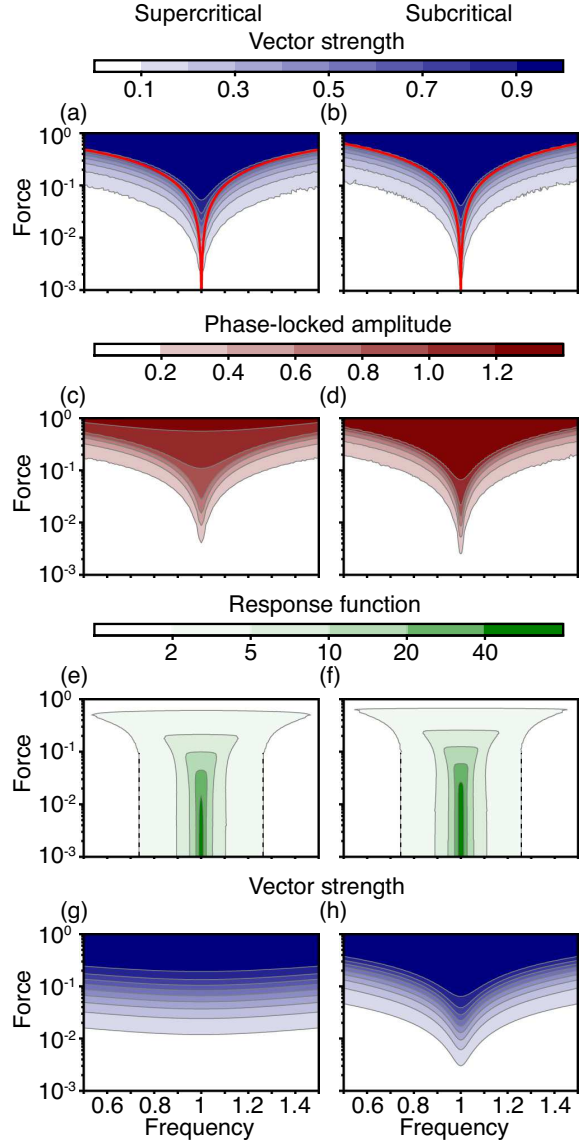}
\caption{State diagrams of the driven stochastic system ($d = 0.01$) as a function of the forcing amplitude and frequency. (a, b) Contours of constant vector strength are shown for $\mu = 1$. The boundaries for perfect phase locking in the deterministic limit are shown (red). (c, d) Contours for which the phase-locked amplitude is constant are shown for $\mu = 1$. (e, f) Contours defined by a constant value for the response function are shown for $\mu = 1$. The $|\tilde{\chi}| = 2$ contours are approximated by lines of constant detuning from $f=0.1$ to $f=0.001$ (black dashed). (g) Contours of constant vector strength are shown for $\mu = -0.5$. (h) Contours of constant vector strength are shown for $\mu = -0.2$, at which there is coexistence between a stable fixed point and a stable limit cycle. The intensity scale for the vector strength above panels a and b also applies to panels g and h. Small irregularities are evident for some contours owing to numerical error.\label{ArnoldTongueStoch_fig}
}
\end{figure}
For the stochastic cases, larger forcing amplitudes and less detuning are required to achieve a threshold level of entrainment as the noise increases (Fig.~\ref{ArnoldTongueStoch_fig}). Unlike the situation for the deterministic limit, entrainment is not perfect on the quiescent sides of the bifurcations. Noise simplifies the state diagrams corresponding to the coexistence region near a subcritical Hopf bifurcation: there is a single entrainment wedge for each value of the stochastic system's vector strength. As for the cases without noise, the entrainment wedges broaden and rise to larger forcing amplitudes as the control parameter decreases.

The state diagrams for the phase-locked amplitude have a structure similar to those corresponding to the vector strength. The phase-locked amplitude increases if either the forcing grows or the detuning falls. In contrast, response-function state diagrams possess contours that change shape as the value of $|\tilde{\chi}|$ declines. For small detuning, raising the forcing amplitude decreases the response function. Sufficiently large detuning, however, creates a peak in the response function at a specific forcing amplitude (Fig.~\ref{chiVSvsf_detuned_fig}), corresponding to an increase in the width of the region enclosed by a contour as the forcing rises.

\section{Discussion\label{discussion}}
Many studies of oscillators employ power spectra, residence-time distributions, and information metrics to demonstrate a system's response to sinusoidal forcing \cite{gammaitoni98,lindner04,pikovsky03}. Here we study two comparatively neglected measures, the phase-locked amplitude and the vector strength, that can be calculated directly from the probability density.

Because the phase-locked amplitude and the vector strength are zero in the absence of deterministic forcing, they allow a system to clearly distinguish sinusoidal forcing from fluctuations owing to noise. In contrast, the average of the amplitude's magnitude $\langle\rho\rangle_\mathrm{s}$ is non-zero in the absence of input. If the goal of a detector is to determine whether a sinusoidal signal is present, the phase-locked amplitude and the vector strength suffice, whereas the average of the amplitude's magnitude does not differentiate between the signal and the noise.

In the deterministic limit, the response of a supercritical Hopf oscillator is qualitatively different from that of a subcritical Hopf oscillator. A supercritical Hopf oscillator exhibits a single stable response to sinusoidal forcing. In contrast, a subcritical Hopf oscillator possesses two stable responses for a range of negative control-parameter values, and the response amplitude changes discontinuously at each end of this range. Bivalued or parametrically discontinuous responses are not desirable properties for a signal detector. In the presence of noise, the subcritical Hopf oscillator responds in a manner qualitatively similar to the supercritical Hopf oscillator: there is a single, averaged response at all values of the control parameter. The response within the deterministically-bivalued region is temporally irregular, however, because noise induces transitions between the stable loci. For this reason, a noisy oscillator may perform better as a detector of sinusoidal signals and be more robust to perturbations in parameter values when it operates near a supercritical Hopf rather than a subcritical Hopf bifurcation.

For simplicity, we specialize to the case in which the resonant frequency is independent of the forcing amplitude ($b' = 0$ and $c' = 0$). When $b' \neq 0$ in the supercritical case and in the absence of forcing, an analytical solution to the Fokker-Planck equation exists, which can be used to demonstrate the non-Lorentzian form of the power spectrum \cite{gleeson06}. For weak forcing and small detuning, an analytical solution to the Fokker-Planck equation permits the calculation of the linear response of a supercritical oscillator \cite{julicher09}. More generally, the Fokker-Planck equation (Eq.~\ref{FProt}) in the instance of nonzero $b'$ and $c'$ can be solved numerically through the strategy that we employ here to solve the equation for detuned forcing. The exact solution that we report (Eq.~\ref{sdist}) can be used to set the initial and boundary conditions. Although we postpone a description of this more general solution to a subsequent publication, we offer some comments here. In the deterministic limit for a self-oscillating supercritical Hopf oscillator, there can exist many stable responses at a single driving frequency when $b'$ is sufficiently large \cite{zhang11,wiser15}. The response is undoubtedly even more complicated in the case of a subcritical Hopf oscillator when $b'$ and $c'$ are nonzero. By definition, however, the averaged response in the presence of noise is single-valued. Nonetheless, the stochastic dynamics of a Hopf oscillator in the most general case likely involves transitions between all stable responses.

In general, the effects of noise are small when $f \gg \sqrt{2d}$ and are great when $f \ll \sqrt{2d}$. The presence of any level of noise, however, changes the behavior of a Hopf oscillator qualitatively. For example, noise creates a peak in the vector strength as a function of the control parameter. To illustrate the effects of noise clearly in many figures, we choose $f \ll \sqrt{2d}$. This choice demonstrates that an oscillator can detect sinusoidal input in the presence of large amounts of noise. Such a regime may correspond to the threshold of hearing, for the sensory detectors of the auditory system apparently function as noisy Hopf oscillators driven by weak periodic forces owing to quiet sounds \cite{salvi15,salvi16}.

A deterministic oscillator responds to sinusoidal forcing in a qualitatively different manner when self-oscillating than when quiescent. A quiescent oscillator is entrained by forcing of any amplitude and frequency, whereas a self-oscillating system is entrained only when the forcing is large and the detuning is small. Any level of noise eliminates this difference. The state diagrams of a forced Hopf oscillator operating on the quiescent side of the bifurcation are qualitatively similar to the diagrams corresponding to the oscillatory side.

Because the phase-locked amplitude at resonance and the sharpness of tuning of a quiescent supercritical Hopf oscillator increase as the system approaches the bifurcation, it has been suggested that sinusoidal-signal detectors, such as those in the auditory system, should operate near and on the quiescent side of a supercritical Hopf bifurcation \cite{choe98,eguiluz00,camalet00}. We find, however, that the phase-locked amplitude at resonance and tuning sharpness continue to grow as the control parameter is raised even after the system crosses such a bifurcation. If the phase-locked amplitude and frequency selectivity of a sinusoidal-signal detector are required to be as great as possible, then an oscillator should be set to spontaneously oscillate far from a Hopf bifurcation. Moreover, noise has the largest effect on the phase-locked amplitude when the system operates near a Hopf bifurcation. These requirements and the effect of noise seem to imply that a detector of sinusoidal signals does not perform best when it operates near a Hopf bifurcation.

Detectors of sinusoidal signals, nevertheless, gain other advantages by operating close to a Hopf bifurcation. In the presence of noise, the phase-locked amplitude and the vector strength corresponding to detuned forcing peak when the system is close to the bifurcation. In addition, the range of stimulus frequencies that can be detected is maximized near the Hopf bifurcation. This frequency bandwidth is most robust to changes in the value of the control parameter on the larger-$\mu$ side of the bandwidth's peak. For sufficiently large noise levels, an oscillator can detect \emph{detuned} sinusoidal stimuli best when it spontaneously oscillates near a Hopf bifurcation.

In the presence of noise and detuning, the phase-locked amplitude, range of compression, degree of entrainment, and threshold bandwidth peak at different values of the control parameter near the critical point. A compromise must consequently be made depending on which features of the signal detector are essential for its function. For example, if very strong compression in a supercritical Hopf oscillator is desired then a spontaneously oscillating system that is operating far from the bifurcation should be chosen \cite{lindner09}. The oscillator's detection bandwidth would be small, however, limiting its response to detuned stimuli. In contrast, if an oscillator is required to detect sinusoidal signals within a specific frequency bandwidth and to be robust to changes in parameter values, then the detector should be set to spontaneously oscillate near a Hopf bifurcation. Under these conditions, the oscillator is sharply tuned---but not too sharply tuned---and displays a wide dynamic range owing to compression of the input amplitude. For stimulus frequencies within the detection bandwidth, the detector's phase-locked amplitude and degree of entrainment are large but are not maximized as functions of the control parameter.

In conclusion, we show that an active, noisy oscillator can detect sinusoidal stimuli best when it \emph{spontaneously oscillates} near a Hopf bifurcation. To detect sinusoidal signals in a noisy environment, operation near criticality has apparently evolved in biological systems \cite{omaoileidigh12}, might be enforced by homeostatic mechanisms \cite{milewski17}, and can be employed in the construction of artificial devices \cite{martignoli07}.

\appendix
\section{Numerical methods \label{numericalFPApp}}
To find the detuned steady-state solution of Eqs.~\ref{FProt} numerically for each set of parameters, initial and boundary conditions must be specified. An injudicious choice for these conditions can result in large computation times and inconsistent results. An exact solution for the probability density of a noisy Hopf oscillator in response to tuned forcing allows us to circumvent this difficulty. Because the tuned solution is close to the steady-state solution for weak detuning, it constitutes a suitable initial condition. Larger computation times are required for greater detuning. Far from the origin $(0, 0)$, the steady-state distribution in the detuned case is close to the tuned distribution. Thus, the tuned distribution can be used to estimate the boundary beyond which the detuned density contributes little to values of the responsiveness measures.

We solve the system for the distribution $P_\mathrm{p}(\rho,\psi,t)$ in polar coordinates with periodic boundary conditions in the phase coordinate $P_\mathrm{p}(\rho,2\pi,t)=P_\mathrm{p}(\rho,0,t)$ and $\partial_\psi P_\mathrm{p}(\rho,\psi,t)|_{\psi=2\pi}=\partial_\psi P_\mathrm{p}(\rho,\psi,t)|_{\psi=0}$. In the radial direction we use the Dirichlet boundary condition $P_\mathrm{p}(\rho_\mathrm{max},\psi,t)=P_\mathrm{ps}(\rho_\mathrm{max},\psi)$, in which $P_\mathrm{ps}(\rho,\psi)= \rho P_\mathrm{s}(y(\rho,\psi))$ is the steady-state solution for zero detuning in polar coordinates. We define the integral
\begin{eqnarray}
I(\rho_\mathrm{min},\rho_\mathrm{max}) &=&\int_{\rho_\mathrm{min}}^{\rho_\mathrm{max}}\,d\rho \rho e^{\frac{\mu\rho^2}{2d}+\frac{b\rho^4}{4d}+\frac{c\rho^6}{6d}} \int_{0}^{2\pi}\,d\psi e^{\frac{f\rho\cos{\psi}}{d}}\mathrm{,}
\label{Ieqn}
\end{eqnarray}
such that $I(0,\infty)^{-1}$ is the normalization constant for $P_\mathrm{ps}(\rho,\psi)$. The maximum value for $\rho$, $\rho_\mathrm{max}$, is selected such that error $1-I(0,\rho_\mathrm{max})/I(0,\infty) < 2\times10^{-3}$. The initial condition is chosen to be $P_\mathrm{p}(\rho,\psi,0)=P_\mathrm{ps}(\rho,\psi)$. 

In polar coordinates Eq.~\ref{FProt} is transformed to an equation with terms that diverge as $\rho \to 0$
\begin{eqnarray}
\label{FPprot}
\partial_\mathrm{t} P_\mathrm{p}
&=&-\partial_\mathrm{\rho}\left[\left(\mu \rho+b\rho^3+c\rho^5+f\cos\psi+\frac{d}{\rho}\right)P_\mathrm{p}-d\partial_\mathrm{\rho}P_\mathrm{p}\right]\nonumber\\
&-&\partial_\mathrm{\psi}\left[\left(-\delta\omega+b'\rho^2+c'\rho^4-\frac{f\sin\psi}{\rho}\right)P_\mathrm{p}-\frac{d}{\rho^2}\partial_\mathrm{\psi}P_\mathrm{p}\right]\mathrm{.}
\end{eqnarray}
To avoid numerical difficulties associated with the divergence we define a minimum value for $\rho$, $\rho_\mathrm{min}$, and use the condition $P_\mathrm{p}(\rho_\mathrm{min},\psi,t)=P_\mathrm{ps}(\rho_\mathrm{min},\psi)$. The minimum value for $\rho$, $\rho_\mathrm{min}$, is selected such that error $1-I(\rho_\mathrm{min},\rho_\mathrm{max})/I(0,\infty) < 6\times10^{-3}$. The distribution is normalized at the end of the integration time $t_\mathrm{max}$ to have unit volume over the annulus defined by $\rho\in[\rho_\mathrm{min},\rho_\mathrm{max}]$ and $\psi\in[0,2\pi]$. The steady-state distribution in Cartesian coordinates is then given by $P_\mathrm{s\delta\omega}(y)=P_\mathrm{p}(\rho(y),\psi(y),t_\mathrm{max})/\rho(y)$, in which $t_\mathrm{max}$ is chosen such that $P_\mathrm{s\delta\omega}(y)$ has converged sufficiently. We note that the corresponding radial Langevin equation in polar coordinates possesses the additional drift term $d/\rho$ in the It\^o case in comparison to the Stratonovich form \cite{gardiner}. In the It\^o case we have
\begin{eqnarray}
\dot{\rho}&=&\mu\rho+b\rho^3+c\rho^5+f\cos{\psi}+\eta_\mathrm{R}\cos{\phi}+\eta_\mathrm{I}\sin{\phi}+\frac{d}{\rho}\nonumber\mathrm{,}\\
\dot{\psi}&=&-\delta\omega+b'\rho^2+c'\rho^4-\frac{f\sin{\psi}}{\rho}-\frac{\eta_\mathrm{R}\sin{\phi}}{\rho}+\frac{\eta_\mathrm{I}\cos{\phi}}{\rho}\mathrm{,}
\label{polarNF}
\end{eqnarray}in which $\eta_\mathrm{R}$ and $\eta_\mathrm{I}$ are respectively the real and imaginary parts of the noise $\eta(t)$. Both pairs of polar Langevin equations, however, correspond to the same deterministic limit and the same Fokker-Planck equation.

Using NDSolve[ ] in Mathematica 10.1--11.2, we solve Equation \ref{FPprot} with the method of lines, which approximates the Fokker-Planck equation as a set of ordinary differential equations to be integrated in time. The value of $t_\mathrm{max}$ is chosen empirically for each data point such that the responsiveness measures changed by less than 1 \% when $t_\mathrm{max}$ is halved.

The imperfectly phase-locked deterministic results of Figs.~\ref{chiVSvsmu_detuned_fig}, \ref{chiVSvsw_fig}, \ref{BWmu_fig}, \ref{chiVSvsf_detuned_fig}, and \ref{ArnoldTongueDet_fig} were obtained from time traces generated by integrating the deterministic limit of Eq.~\ref{HopfNF}. The amplitude of the steady-state response is determined from the last half of each time trace using the finite time discrete Fourier transform, implemented as Fourier[ ] in Mathematica 10.1--11.2. The trace was deemed to have reached steady state if the Fourier amplitude at the driving frequency changed by less than 1 \% upon halving the integration time.

\section{Phase-locked amplitude and linear response\label{chimagApp}}
In the steady state, the average Fourier amplitude in the nonrotating frame is
\begin{eqnarray}
\langle\tilde{z}(\omega')\rangle_\mathrm{zs}&=&\left\langle\frac{1}{T}\int_{-\frac{T}{2}}^\frac{T}{2} e^{-i\omega't}z(t)\,dt\right\rangle_\mathrm{zs}\nonumber\\
&=&\frac{1}{T}\int_{-\infty}^\infty\int_{-\infty}^\infty dz_\mathrm{R}dz_\mathrm{I}\int_{-\frac{T}{2}}^\frac{T}{2} e^{-i\omega't}P_\mathrm{zs}(z,t)z\,dt\nonumber\\
&=&\frac{1}{T}\int_{-\infty}^\infty\int_{-\infty}^\infty dy_\mathrm{R}dy_\mathrm{I}\int_{-\frac{T}{2}}^\frac{T}{2} e^{-i(\omega'-\omega)t}P_\mathrm{s}(y)ye^{i\theta}\,dt\nonumber\\
&=&\frac{1}{T}\int_{-\frac{T}{2}}^\frac{T}{2} e^{-i(\omega'-\omega)t}\langle y\rangle_\mathrm{s}e^{i\theta}\,dt\nonumber\\
&=&\frac{\sin{((\omega'-\omega)T/2)}}{(\omega'-\omega)T/2}\langle y\rangle_\mathrm{s}e^{i\theta}\mathrm{,}
\label{aveFourampl}
\end{eqnarray}
which is either $\langle y\rangle_\mathrm{s}e^{i\theta}$ in the limit $\omega' \to \omega$ or 0 if $\omega' \neq \omega$ and $T \to \infty$. $P_\mathrm{zs}(z,t)$ is the distribution of $z$ in the long-time limit, $\langle \rangle_\mathrm{zs}$ is the corresponding ensemble average, $y = y_\mathrm{R}+ iy_\mathrm{I}$, $\theta$ is the phase of the stimulus force, and $\langle y\rangle_\mathrm{s}$ is independent of time. We note that Eq.~\ref{aveFourampl} also holds in the non-isochronous case when stationary solutions of Eq.~\ref{FProt} exist.

From Eqs.~\ref{sdist} and \ref{aveFourampl}, the magnitude of the response $\langle y\rangle_\mathrm{sl}$ for a weak tuned stimulus is given by
\begin{eqnarray}
\left|\langle y\rangle_\mathrm{sl}\right| &\equiv& \left|\iint_{-\infty}^\infty y P_\mathrm{s}\,dy_\mathrm{R}\,dy_\mathrm{I}\right|\nonumber\\
&=& \left|\iint_{-\infty}^\infty y Ne^{\frac{\mu\rho^2}{2d}+\frac{b\rho^4}{4d}+\frac{c\rho^6}{6d}+\frac{fy_\mathrm{R}}{d}}\,dy_\mathrm{R}\,dy_\mathrm{I}\right|\nonumber\\
&\approx& \left|\iint_{-\infty}^\infty y P_\mathrm{s0} \left(1+\left(\frac{y_\mathrm{R}}{d}+\frac{(\partial_\mathrm{f}N)_\mathrm{f=0}}{N_0}\right)f\right)\,dy_\mathrm{R}\,dy_\mathrm{I}\right|\nonumber\\
&=&\frac{\left|\langle y_\mathrm{R}y\rangle_\mathrm{s0}\right|f}{d}\mathrm{,}
\label{phaselocklin}
\end{eqnarray}
in which $P_\mathrm{s0}$ is the steady-state distribution in the absence of forcing and detuning, $N_0$ is the normalization constant for this distribution, and we have expanded the distribution $P_\mathrm{s}$ to linear order in $f$ on the third line. The linear phase-locked response (Eq.~\ref{chilin}) follows directly.

The same approximation for $P_\mathrm{s}$ yields the vector strength to linear order in $f$
\begin{eqnarray}
V_\mathrm{l} &=& \left|\iint_{-\infty}^\infty e^{i\psi} P_\mathrm{s}\,dy_\mathrm{R}\,dy_\mathrm{I}\right|\nonumber\\
&\approx& \left|\iint_{-\infty}^\infty e^{i\psi} P_\mathrm{s0} \left(1+\left(\frac{y_\mathrm{R}}{d}+\frac{(\partial_\mathrm{f}N)_\mathrm{f=0}}{N_0}\right)f\right)\,dy_\mathrm{R}\,dy_\mathrm{I}\right|\nonumber\\
&=&\frac{\left|\langle y_\mathrm{R}e^{i\psi}\rangle_\mathrm{s0}\right|f}{d}\mathrm{.}
\end{eqnarray}

According to Eq.~\ref{phaselocklin}, the linear response is given by
\begin{eqnarray}
\left|\langle y\rangle_\mathrm{sl}\right| &=&\frac{\left|\langle y_\mathrm{R}y\rangle_\mathrm{s0}\right|f}{d}\nonumber\\
&=&\left|\int_{0}^\infty\,d\rho\int_{0}^{2\pi}\,d\psi\rho^2(\cos^2{\psi}+i\cos{\psi}\sin{\psi})P_\mathrm{p0}\right|\frac{f}{d}\nonumber\\
&=&\left|\int_{0}^\infty\,d\rho\int_{0}^{2\pi}\,d\psi\rho^2P_\mathrm{p0}\right|\frac{f}{2d}\nonumber\\
&=&\frac{\langle \rho^2\rangle_\mathrm{s0}f}{2d}\mathrm{,}
\label{linrespolar}
\end{eqnarray}
because $P_\mathrm{p0} = \rho P_\mathrm{s0}$ is independent of $\psi$. This expression is the same as the linear, phase-locked amplitude found previously for a supercritical Hopf bifurcation \cite{julicher09}. 

The response $\langle y\rangle_\mathrm{s}$ to a tuned stimulus of any amplitude $f$ can also be written as 
\begin{eqnarray}
\langle y\rangle_\mathrm{s} &=&\int_{0}^\infty\,d\rho\int_{0}^{2\pi}\,d\psi yP_\mathrm{ps}\nonumber\\
&=&\frac{N}{N_0}\int_{0}^\infty\,d\rho P_\mathrm{p0}\rho\int_{0}^{2\pi}\,d\psi(\cos{\psi}+i\sin{\psi})e^\frac{f\rho\cos{\psi}}{d}\nonumber\\
&=&\frac{N}{N_0}\int_{0}^\infty\,d\rho P_\mathrm{p0}\rho 2\pi I_\mathrm{1}\left(\frac{f\rho}{d}\right)\nonumber\\
&=&\frac{\left\langle \rho I_\mathrm{1} \left(\frac{f\rho}{d}\right) \right\rangle_\mathrm{s0}}{\left\langle I_\mathrm{0} \left(\frac{f\rho}{d}\right) \right\rangle_\mathrm{s0}}\mathrm{,}
\label{phaselockBessel}
\end{eqnarray}
in which the normalization constant $N$ is given by
\begin{eqnarray}
N^{-1} &=&\int_{0}^\infty\,d\rho \rho e^{\frac{\mu\rho^2}{2d}+\frac{b\rho^4}{4d}+\frac{c\rho^6}{6d}} \int_{0}^{2\pi}\,d\psi e^{\frac{f\rho\cos{\psi}}{d}}\nonumber\\
&=&N_\mathrm{0}^{-1}\int_{0}^\infty\,d\rho P_\mathrm{p0} 2\pi I_\mathrm{0}\left(\frac{f\rho}{d}\right)\nonumber\\
&=&N_\mathrm{0}^{-1}\left\langle I_\mathrm{0} \left(\frac{f\rho}{d}\right) \right\rangle_\mathrm{s0}\mathrm{,}
\label{Neqn}
\end{eqnarray}
and
\begin{equation}
I_\mathrm{n}(x) \equiv \frac{1}{\pi}\int_{0}^{\pi}\,d\theta e^{x\cos{\theta}}\cos{n\theta}
\end{equation}
is an expression for the modified Bessel function of the first kind of integer order $n$.

If the noise level is sufficiently small, then $\rho$ varies little and can be replaced by a deterministic estimate $\rho_\mathrm{d}(f)$ \cite{lindner09}. From Eq.~\ref{phaselockBessel} the response function's magnitude is then approximately
\begin{equation}
|\tilde{\chi}| = \frac{\rho_\mathrm{d}(f)}{f}\left|\frac{I_\mathrm{1} \left(\frac{f\rho_\mathrm{d}(f)}{d}\right)}{I_\mathrm{0} \left(\frac{f\rho_\mathrm{d}(f)}{d}\right) }\right|\mathrm{.}
\end{equation}
This expression has been found previously through a different approach and yields the linear response function $|\tilde{\chi}_\mathrm{l}| = \frac{\rho_\mathrm{d}^2(0)}{2d}$ in the limit $f \to 0$ in concordance with Eq.~\ref{linrespolar} \cite{lindner09}.

Similarly to Eq.~\ref{phaselockBessel}, it is straightforward to show that
\begin{eqnarray}
\langle e^{i\psi}\rangle_\mathrm{s} &=&\frac{\left\langle I_\mathrm{1} \left(\frac{f\rho}{d}\right) \right\rangle_\mathrm{s0}}{\left\langle I_\mathrm{0} \left(\frac{f\rho}{d}\right) \right\rangle_\mathrm{s0}}\mathrm{.}
\label{VSBessel}
\end{eqnarray}
Equation \ref{PLvVSBessel} follows immediately.

\section{Polar amplitude moments\label{amplmomApp}}
The amplitude moments of the polar distribution $P_\mathrm{p}(\rho,\psi,t)$ are defined by
\begin{equation}
\langle \rho^\mathrm{n}(t)\rangle \equiv \int_{0}^\infty\,d\rho\int_{0}^{2\pi}\,d\psi\rho^\mathrm{n}P_\mathrm{p}(\rho,\psi,t)\mathrm{.}
\end{equation}
In the steady state, the even and odd moments can be expressed as respectively
\begin{eqnarray}
\langle \rho^\mathrm{2n}\rangle_\mathrm{s} &=& N (2d)^\mathrm{n} \frac{\partial^\mathrm{n}}{\partial \mu^\mathrm{n}} N^{-1}\nonumber\\
&\mathrm{and}&\nonumber\\
\langle \rho^\mathrm{2n-1}\rangle_\mathrm{s} &=& N (2d)^\mathrm{n} \frac{\partial^\mathrm{n}}{\partial \mu^\mathrm{n}} M\mathrm{,}
\end{eqnarray}
in which $n$ is a positive integer, $N$ is the normalization constant (Eq.~\ref{Neqn}), and
\begin{equation}
M \equiv \int_{0}^\infty\,d\rho e^{\frac{\mu\rho^2}{2d}+\frac{b\rho^4}{4d}+\frac{c\rho^6}{6d}} \int_{0}^{2\pi}\,d\psi e^{\frac{f\rho\cos{\psi}}{d}}\mathrm{.}
\end{equation}
The constants $N$ and $M$ can be expressed in terms of the zeroth order modified Bessel function of the first kind $I_{0}(x)$, yielding
\begin{eqnarray}
\langle \rho^\mathrm{2n}\rangle_\mathrm{s} &=& \frac{N_0 (2d)^\mathrm{n}}{\left\langle I_\mathrm{0} \left(\frac{f\rho}{d}\right) \right\rangle_\mathrm{s0}} \frac{\partial^\mathrm{n}}{\partial \mu^\mathrm{n}} \left(\frac{\left\langle I_\mathrm{0} \left(\frac{f\rho}{d}\right) \right\rangle_\mathrm{s0}}{N_0} \right)\nonumber\\
&\mathrm{and}&\nonumber\\
\langle \rho^\mathrm{2n-1}\rangle_\mathrm{s} &=& \frac{N_0 (2d)^\mathrm{n}}{\left\langle I_\mathrm{0} \left(\frac{f\rho}{d}\right) \right\rangle_\mathrm{s0}} \frac{\partial^\mathrm{n}}{\partial \mu^\mathrm{n}} \left(\frac{\left\langle \frac{I_\mathrm{0} \left(\frac{f\rho}{d}\right)}{\rho} \right\rangle_\mathrm{s0}}{N_0} \right)\mathrm{,}
\end{eqnarray}
in which $n$ is a positive integer. These expressions show how amplitude moments of the unforced and forced Hopf oscillator can be calculated systematically from knowledge of only a normalization constant $N_0$ and two expectation values.

In the absence of forcing,
\begin{eqnarray}
\langle \rho^\mathrm{2n}\rangle_\mathrm{s0} &=& N_0 (2d)^\mathrm{n} \frac{\partial^\mathrm{n}}{\partial \mu^\mathrm{n}} N_0^{-1} \nonumber\\
&\mathrm{and}&\nonumber\\
\langle \rho^\mathrm{2n-1}\rangle_\mathrm{s0} &=& N_0 (2d)^\mathrm{n} \frac{\partial^\mathrm{n}}{\partial \mu^\mathrm{n}} \left(\frac{\left\langle \rho^{-1} \right\rangle_\mathrm{s0}}{N_0} \right)\mathrm{.}
\end{eqnarray}
In the supercritical case, the normalization constant $N_0$ is given by
\begin{equation}
N_0^{-1} = \frac{de^{-\frac{\mu^2}{4bd}}\pi^\frac{3}{2}\left(1+\mathrm{erf}\left(\frac{\mu}{2\sqrt{-bd}}\right)\right)}{\sqrt{-bd}}\mathrm{,}
\end{equation}
in which $\mathrm{erf}$ is the error function. The amplitude moments can then be derived analytically in terms of Bessel functions and error functions.

\section{Deterministic limit\label{perfectlockApp}}
In polar coordinates $y=\rho e^{i\psi}$ and Eq.~\ref{HopfNFrot} becomes
\begin{eqnarray}
\dot{\rho}&=&\mu\rho+b\rho^3+c\rho^5+f\cos{\psi}\nonumber\mathrm{,}\\
\dot{\psi}&=&-\delta\omega+b'\rho^2+c'\rho^4-\frac{f\sin{\psi}}{\rho}
\label{polarNF}
\end{eqnarray}
in the deterministic limit. The fixed points $\rho^*$ and $\psi^*$ then satisfy
\begin{eqnarray}
(\mu+b\rho^{*2}+c\rho^{*4})^2\rho^{*2}+(-\delta\omega+b'\rho^{*2}+c'\rho^{*4})^2\rho^{*2}&=&f^2\nonumber\\
-\delta\omega+b'\rho^{*2}+c'\rho^{*4}-\frac{f\sin{\psi^*}}{\rho^*}&=&0\mathrm{,}
\label{perfectlock_eq}
\end{eqnarray}
and their linear stability is determined by the eigenvalues of the Jacobian
\begin{equation}
\mathbf{J}\equiv\left( \begin{array}{cc} \mu+3b\rho^{*2}+5c\rho^{*4} & \delta\omega\rho^{*}-b'\rho^{*3}-c'\rho^{*5} \\
-\frac{\delta\omega}{\rho^{*}}+3b'\rho^{*}+5c'\rho^{*3} & \mu+b\rho^{*2}+c\rho^{*4} \end{array} \right)\mathrm{.}
\label{Jac_eq}
\end{equation}
Lines of saddle node bifurcations are defined by $\mathrm{Det}(\mathbf{J})  = 0$, whereas lines of Hopf bifurcations are given by $\mathrm{Tr}(\mathbf{J})=0$ and $\mathrm{Det}(\mathbf{J})  > 0$, in which $\mathrm{Tr}(\mathbf{J})$ and $\mathrm{Det}(\mathbf{J})$ are respectively the trace and determinant of $\mathbf{J}$. The Hopf bifurcation lines end at Takens-Bogdanov points defined by $\mathrm{Tr}(\mathbf{J})=0$ and $\mathrm{Det}(\mathbf{J})  = 0$ \cite{kuznetsov}. Two saddle-node bifurcation lines can meet tangentially at cusp bifurcations. In Fig.~\ref{ArnoldTongueDet_fig}, the full extents of the bifurcation lines are not shown, Takens Bogadanov points are not denoted, cusp points are not illustrated, and global bifurcations are not depicted.
 
In the isochronous case $b' = c' = 0$ and for tuned forcing $\delta\omega = 0$, Eq.~\ref{perfectlock_eq} becomes
\begin{equation}
(\mu+by_\mathrm{R}^{*2}+cy_\mathrm{R}^{*4})^2y_\mathrm{R}^{*2}=f^2\mathrm{,}
\label{ischron_perfectlock_eq}
\end{equation}
in which $\psi^* = 0$ or $\pi$ such that $y_\mathrm{I}^* = 0$. Values of $y_\mathrm{R}^{*}$ are given by real solutions of the saddle-node bifurcation condition, which simplifies to $\mu+3by_\mathrm{R}^{*2}+5cy_\mathrm{R}^{*4} = 0$.

In the tuned supercritical case, Eq.~\ref{ischron_perfectlock_eq} defines three fixed points for $f < \sqrt{-4\mu^3/27b}$ when $\mu > 0$, a stable fixed point, an unstable fixed point, and a saddle point, whereas a single stable fixed point exists for larger forces. For $\mu > 0$ or $\mu < 9 b^2/20 c$, the resonantly forced subcritical and supercritical oscillators exhibit similar phenomenology. In the range $9 b^2/20 c < \mu < 0$, however, the subcritical Hopf oscillator has two stable responses for tuned sinusoidal forcing satisfying $f_\mathrm{1}(\mu,b,c) < f < f_\mathrm{2}(\mu,b,c)$, in which
\begin{eqnarray}
f_\mathrm{1}(\mu,b,c) &=& \sqrt{\frac{-3b-\sqrt{9b^2-20c\mu}}{10c}}\left(\frac{3b^2-20c\mu+ b\sqrt{9b^2-20c\mu}}{25c}\right)\mathrm{, if}\,\mu < \frac{b^2}{4c}\mathrm{, and}\nonumber\\
f_\mathrm{2}(\mu,b,c) &=& \sqrt{\frac{-3b+\sqrt{9b^2-20c\mu}}{10c}}\left(\frac{3b^2-20c\mu- b\sqrt{9b^2-20c\mu}}{25c}\right)\mathrm{.}
\end{eqnarray}

\section{The limits of large-magnitude control parameters \label{CPlimApp}}
Let $\mu_\mathrm{s} < 0$ denote the value of the control parameter below which a single stable solution of Eq.~\ref{perfectlock_eq} exists. In the absence of noise and when $\mu < \mu_\mathrm{s}$, the system is perfectly phase-locked to the stimulus such that $R=\rho^*$, in which $\rho^*$ is the value of $\rho$ at the fixed point. To satisfy Eq.~\ref{perfectlock_eq} as $\mu \to -\infty$ for finite $\rho^*$, $\rho^* \to 0$. In the noisy case, $P_\mathrm{s0}$ becomes sharply peaked at $(0, 0)$ as $\mu \to -\infty$. Consequently, $\langle I_\mathrm{n}(f\rho/d)\rangle_\mathrm{s0} \to I_\mathrm{n}(0)$ and $\langle\rho I_\mathrm{1}(f\rho/d)\rangle_\mathrm{s0} \to \rho I_\mathrm{1}(0) = 0$. Equation \ref{phaselockBessel} then yields $\langle y \rangle_\mathrm{s} \to 0$, as $I_\mathrm{0}(0) = 1$ is finite. Thus $R \to 0$ as $\mu \to -\infty$ with or without noise.

In the presence of noise, $V < V_{\delta\omega=0}$, in which $V_{\delta\omega=0}$ is the vector strength corresponding to a tuned stimulus. According to Eq.~\ref{VSBessel}, $V_{\delta\omega=0} \to I_\mathrm{1}(0)/I_\mathrm{0}(0) = 0$ as $\mu \to -\infty$. Thus noise causes $V \to 0$ as $\mu \to -\infty$ for both tuned and detuned stimuli.

The system becomes essentially deterministic as $\mu \to \infty$, but the detuned response $z = \rho e^{i\phi}$ is not phase-locked to the stimulus. In this limit, forcing has a negligible effect on the amplitude $\rho$, which is approximately given by the first line of Eq.~\ref{polarNF} with $\dot{\rho}=0$ and $f=0$. In other words, as $\mu \to \infty$, the supercritical amplitude $\rho^* \sim \mu^\frac{1}{2}$ and the subcritical amplitude $\rho^* \sim \mu^\frac{1}{4}$, and thus $\rho^* \to \infty$. Because $\tan\phi = z_\mathrm{I}/z_\mathrm{R}$, the deterministic limit of Eq.~\ref{HopfNF} then yields the dynamics of $\phi$ given by
\begin{eqnarray}
\dot{\phi} &=& \omega_0 + b'\rho^{*2} + c'\rho^{*4} + f\frac{\sin{(\omega t + \theta)}}{\rho^{*}}\nonumber\\
&\approx& \omega_0 + b'\rho^{*2} + c'\rho^{*4}\mathrm{\ for\ }\rho^*\gg \frac{f}{\omega_0}\mathrm{.}
\end{eqnarray}
Thus $\phi \approx (\omega_0 + b'\rho^{*2} + c'\rho^{*4})t + \phi_0 \equiv \Omega t+\phi_0$, in which $\phi_0$ is a constant. The response is now given by $z \approx \rho^*e^{i(\Omega t+\phi_0)}$, for which the Fourier amplitude is $|\tilde{z}(\omega')| \approx \rho^*\frac{\sin{((\omega'-\Omega)T/2)}}{(\omega'-\Omega)T/2}$. The phase-locked amplitude at the frequency of driving $\omega$ approaches zero if $\omega \neq \Omega$ and $T \to \infty$, in which $\Omega = \omega_0$ when $b' = 0$ and $c'=0$.

As $\mu\to\infty$, the deterministic vector strength can be calculated from Eq.~\ref{VS} through the expression for $\phi$ calculated above. The phase difference between the response and the driving is then given by $\psi \approx (\Omega-\omega) t+\phi_0-\theta$. When $b' = 0$ and $c'=0$, then $\Omega \approx \omega_0$, yielding $V \to 0$ for a detuned stimulus. In the limit $\mu\to\infty$, the stochastic system's detuned vector strength is less than the deterministic case, such that $V \to 0$ with or without noise.

The phase-locked amplitude and the stochastic system's vector strength always peak at a finite value of the control parameter, because $R \to 0$ and $V \to 0$ as the control parameter $\mu \to \pm \infty$ and there is a finite value of $\mu$ for which these quantities are positive.

\section*{Acknowledgements}
We thank the members of our research group for constructive comments on the manuscript. A.~J.~H.~is an Investigator of Howard Hughes Medical Institute.
\newpage
\bibliography{ear.bib}
\end{document}